Pressure-Induced Effects on the Structure of the FeSe Superconductor

Jasmine N. Millican[1*], Daniel Phelan[1], Evan L. Thomas[2], Juscelino B. Leão[1], and Elisabeth Carpenter[1]

[1]*NIST Center for Neutron Research, National Institute of Standards and Technology, 100 Bureau Dr. MS 6102, Gaithersburg, MD 20899-6102, USA.*
[2]*Ceramics Division, National Institute of Standards and Technology, 100 Bureau Dr. MS 8520, Gaithersburg, MD 20899-8520, USA.*

*To whom correspondence should be addressed

E-mail: jasmine.millican@nist.gov

Telephone: (301) 975-6422

Fax: (301) 921-9847

**Abstract**

A polycrystalline sample of FeSe, which adopts the tetragonal PbO-type structure ($P4/nmm$) at room temperature, has been prepared using solid state reaction. We have investigated pressure-induced structural changes in tetragonal FeSe at varying hydrostatic pressures up to 0.6 GPa in the orthorhombic ($T = 50$ K) and tetragonal ($T = 190$ K) phases using high resolution neutron powder diffraction. We report that the structure is quite compressible with a bulk modulus $\approx 31$ GPa to 33 GPa and that the pressure response is anisotropic with a larger compressibility along the $c$-axis. Key bond angles of the SeFe$_4$ pyramids and FeSe$_4$ tetrahedra are also determined as a function of pressure.

# 1. Introduction

The emergence of superconductivity in the Fe-based oxypnictides has led to interest in new Fe-based materials, which may also exhibit superconductivity [1]. One of the particularly interesting aspects of superconductivity in these oxypnictides is that a strong relationship between the superconducting state and pressure (either chemical or externally applied) has been reported in a number of materials. The effect of chemical pressure is evident in the 1:1:1:1 compounds (the $(RE)FeAsO_{1-x}$ series, where $RE$ is a Rare Earth atom), where $T_c$ is enhanced by using small $RE$ atoms [2]. As for externally applied pressure, in the $LaFeAsO_{1-x}F_x$ series a sharp increase in $T_c$ is observed at modestly applied pressures, followed by a slow decline in the value of $T_c$ above an optimal pressure ($\approx$ 4 GPa for $LaFeAsO_{0.89}F_{0.11}$) [3]. Even more dramatic observations have been noted for the 1:2:2 compounds. $CaFe_2As_2$ [4, 5], $SrFe_2As_2$ [6], and $BaFe_2As_2$ [6] all exhibit pressure-induced transitions from a non-superconducting state at ambient pressure to a superconducting state under externally applied pressure. On the other hand, in the 1:1:1 compound, LiFeAs, externally applied pressure slowly reduces $T_c$ [7].

Recently, superconductivity has also been discovered in FeSe ($T_c$ = 8 K) [8], which has a tetragonal Pb-O type structure ($P_4/nmm$) at room temperature and planar FeSe slabs that resemble the FeAs slab layers in the $REFeAsO_{1-x}F_x$ compounds [9]. Interestingly, the mechanism of superconductivity in FeSe appears to be unconventional [10] and is currently being scrutinized. Recently an enormous effect of applied pressure on the $T_c$ of FeSe was reported [11]: the onset temperature of superconductivity was shown to increase at 9.1 K/GPa, which is a larger rate of increase than for any of the Fe-As compounds in the superconducting state. However, currently there is lacking information as to how the crystal structure changes under applied pressure. In this manuscript, we report the effect of externally applied hydrostatic pressure on the structure of FeSe at 50 K and 190 K as determined by using high resolution neutron powder neutron diffraction with varying pressures up to 0.6 GPa. Structural parameters (lattice constants, unit cell volumes, and atomic positions) as well as important bond angles and distances are determined as a function of pressure.

## 2. Experimental

Polycrystalline FeSe was synthesized as reported in Ref. [12] with intitial starting materials of Fe and Se in the molar ratio of 1: 0.82. The previous refinements indicated that the final composition of sample synthesized in this manner is nearly stoichiometric FeSe, which is consistent with previously reported findings [13]. A sample volume of approximately 1.5 cm$^3$ was loaded into an Al 7075-T6 alloy pressure cell, which was connected to an external piston driven pressure intensifier, and pressurized using helium as a pressure medium [14]. The pressure cell was loaded into a top-loading Closed Cycle Refrigerator (CCR) for temperature control. Pressure homogeneity across the sample was ensured by keeping all measurements above the melting curve of the *P* vs. *T* of helium.

Neutron powder diffraction data were collected using the BT-1 32 detector high-resolution neutron powder diffractometer at the NIST Center for Neutron Research (NCNR). A Ge(311) monochromator with a 75° takeoff angle, $\lambda$ = 2.0785 Å, and in-pile collimation of 152 of arc was used. Data were collected over the 2$\theta$ range of 3° to 166° with a step size of 0.05° at various pressures and temperatures. The results reported here were measured upon decreasing pressure. Rietveld refinement of the structure of FeSe was performed using the GSAS software package with the EXPGUI interface [15, 16]. The background was fit using a 13-term shifted Chebychev polynomial. Aluminum peaks, which occur as a result of the pressure vessel, were excluded from the observed pattern. Fe$_7$Se$_8$ impurity peaks were also identified in the pattern, but several small peaks resulting from unknown impurities could not be indexed. The lattice parameters, zero point, scale factor, atomic positions, peak profile (type 3) and isotropic thermal parameters were also refined. A representative diffraction pattern along with the fit from the Rietveld refinement is shown in Fig. 1 for FeSe at 0.6 GPa and 50 K, from which it is apparent that there is a reasonable fit between measured and calculated patterns.

## 3. Results and Discussion

Previously, it was shown that the FeSe undergoes a tetragonal (*P4/nmm*) to orthorhombic phase transition (*Cmme*) below 80 K [9, 12]. Similarly, we observed an orthorhombic pattern at 50 K and a tetragonal pattern at 190 K under ambient pressure in the current experiment. This distortion is evidenced by the splitting of the tetragonal (220) and (221) peaks into the orthorhombic (040), (400), (041), and (401) peaks in the measured neutron diffraction pattern. Fig. 2 shows the evolution of these peaks at $T = 50$ K and $T = 190$ K under various applied pressures. At 190 K, the structure remains tetragonal at all pressures and the application of pressure does not induce a phase transition. Likewise, at 50 K, the structure remains orthorhombic up to 0.6 GPa. It is also clear, however, that the splitting of the (220) and (221) peaks becomes less and less evident as the pressure is increased, which suggests that above some threshold pressure outside of the limits of the current experiment the structure may eventually become the higher symmetry, tetragonal phase. The results of refinements and structural parameters at several selected pressures are shown for both 50 K and 190 K in Table 1.
The pressure dependence of the lattice constants is shown in Fig. 3 at 50 K (a) and 190 K (b). Likewise, the pressure dependence of the unit cell volume is shown in Fig. 3 at 50 K (c) and 190 K (d). The isothermal compressibility, $K = - (1/V)(dV/dP)_T$ and bulk modulus, $B = 1/K$, were determined from the linear fits of the volume versus pressure, as shown in Fig. 3 (c-d), and are provided in Table 2 at 50 K and 190 K. Note that at these modest pressures, fits are relatively insensitive to an increase of the bulk modulus with pressure, so a linear approximation of volume versus pressure is appropriate.

Notably, the bulk modulus of 31 GPa to 33 GPa is significantly smaller than the isothermal bulk modulus ($B$) of 66 GPa, which was reported for optimally doped LaFeAsO$_{0.89}$F$_{0.11}$ [17] and a bulk modulus of 102 GPa, which was reported for NdFeAsO$_{0.88}$F$_{0.12}$ [18]. Thus, when compared to the 1:1:1:1 oxypnictide superconductors, the structure of FeSe is significantly more compressible under modest externally applied pressures. On the other hand, the effect of pressure is not as dramatic as observed in the 1:2:2 compound, CaFe$_2$As$_2$, where the system is driven from an orthorhombic to a 'collapsed tetragonal phase' under only moderate pressure [19].

The linear compressibilities, $K_a$, $K_b$, and $K_c$, where $K_a = -(1/a)(da/dP)_T$, were calculated along the $a$, $b$, and $c$ axes, respectively, from the linear fits of the lattice constants and are provided in Table 2. It is evident that the structure is significantly more compressible along the c-axis, which is a result of the layering along that axis. Although the values of compressibility are smaller in the 1:1:1:1 compounds, similar anisotropy has also been observed [17].

The structure of FeSe is shown in Fig. 4 (a), in which the FeSe$_4$ tetrahedra are highlighted and an individual FeSe$_4$ tetrahedron and SeFe$_4$ pyramid are shown. The nearest neighbor Fe-Fe bond distances at 50 K are shown as a function of pressure in Fig. 4 (b). Contraction of the bond distances occurs in a continuous fashion as a function of increased applied pressure. However, no apparent change was observed for the Fe-Se bond distances, which are not shown here, as pressure was increased. The Se-Fe-Se and Fe-Se-Fe bond angles of the tetrahedra and octahedra at 50 K are shown as a function of pressure in Fig. 4 (c) and (d), respectively. Clearly the FeSe$_4$ tetrahedra are quite distorted from perfect tetrahedra as the Fe-Se-Fe bond angle 1 is significantly less than the ideal angle of 109.47°. As the pressure increases, the tetrahedra actually become even more distorted as the deviation of angle 1 with 109.47° becomes even larger. In this sense, the enhanced superconductivity under applied pressure is accompanied by a distortion of the Fe-Se-Fe tetrahedra. This is a point in which FeSe appears to differ from the 1:1:1:1 and 1:2:2 oxypnictide compounds. Firstly, in the oxypnictides the Fe-As/P-Fe bond angles are actually typically larger than 109.47°. Secondly, in the oxypnictides, $T_c$ increases as the Fe(As/P)$_4$ tetrahedra become less distorted [20], which is not consistent with the present results for FeSe.

## 4. Summary

The structure of FeSe was determined as a function of externally applied hydrostatic pressure up to 0.6 GPa by high resolution neutron diffraction. No pressure- induced phase transition was observed at 50 K or 190 K, but the orthorhombic splitting of Bragg reflections became systematically less evident as the pressure was increased. A significantly larger compressibility (smaller bulk modulus) was found for FeSe than for the 1:1:1:1 type oxypnictide superconductors indicating that FeSe is a softer material, which may serve to explain why the reported value of $dT_c/dP$ [11] is larger. A significantly larger compression under applied pressure was observed along the *c*- axis than along the *a* or *b* axes due to the layering of the crystal structure.

## Acknowledgements

The authors would like to thank J. K. Stalick for insightful discussions and assistance.

**Figure Captions:**

**Figure 1.** Neutron diffraction pattern at T = 50 K and P = 0.6 GPa. Measured intensity is shown in red (+) and calculated pattern is shown in green. The difference curve is shown in black and reflections corresponding to the *Cmme* phase are marked. The missing data correspond to strong aluminum reflections from the pressure cell.

**Figure 2.** Pressure dependence of several selected Bragg reflections in (a) the orthorhombic phase at $T = 50$ K and (b) the tetragonal phase at $T = 190$ K. The intensity at $T = 50$ K, $P = 0.03$ GPa has been multiplied by a constant for easier comparison with the other data.

**Figure 3.** Pressure dependence of the lattice constants in (a) the orthorhombic phase at $T = 50$ K and (b) the tetragonal phase at $T = 190$ K. Pressure dependence of the unit cell volume is shown in (c) and (d) at $T = 50$ K and $T = 190$ K, respectively. Error bars in all figures represent one standard deviation (+/-).

**Figure 4.** (a) The pseudo-layered structure of FeSe in the orthorhombic phase. Fe-Se bonds are highlighted as well as the $FeSe_4$ tetrahedra. On the left side, a lone $SeFe_4$ pyramid (top) and a lone $FeSe_4$ tetrahedron are shown. (b) The two closest Fe-Fe distances. Distance 1 corresponds to the distance between the two Fe atoms forming the Fe-Se-Fe angle 2, and Distance 2 to the distance between the two Fe atoms forming Fe-Se-Fe angle 3 in the $SeFe_4$ pyramid shown in (a). (c) The Se-Fe-Se angles of the $FeSe_4$ tetrahedra as labeled in (a). (d) The Fe-Se-Fe angles of the $SeFe_4$ pyramids as labeled in (a). Structure in (a) was drawn using Balls and Sticks [21].

# References

[1]  Y. Kamihara, T. Watanabe, M. Hirano, H. Hosono, J. Am. Chem. Soc. 130 (2008) 3296-3297.

[2]  X. H. Chen, T. Wu, G. Wu, R. H. Liu, H. Chen, D. F. Fang, Nature 453 (2008) 761-762.

[3]  H. Takahashi, K. Igawa, K. Arii, Y. Kamihara, M. Hirano, H. Hosono, Nature 453 (2008) 376-378.

[4]  T. Park, E. Park, H. Lee, T. Klimczuk, E. D. Bauer, F. Ronning, J. D. Thompson, J. Phys.: Condens. Matter 20 (2008) 322204.

[5]  M. S. Torikachvili, S. L. Bud'ko, N. Ni, P. C. Canfield, Phys. Rev. Lett. 101 (2008) 057006.

[6]  P. L. Alireza, Y. T. C. Ko, J. Gillett, C. M. Petrone, J. M. Cole, G. G. Lonzarich, S. E. Sebastian, J. Phys.: Condens. Matter 21 (2009) 012208.

[7]  M. Gooch, B. Lv, J. H. Tapp, Z. Tang, B. Lorenz, Europhys Lett. 85 (2009) 27005.

[8]  F.-C. Hsu, J. Y. Luo, K. W. Yeh, T. K. Chen, T. W. Huang, P. M. Wu, Y. C. Lee, Y. L. Huang, Y. Y. Chu, D. C. Yan, M. K. Wu, Proc. Natl. Acad. Sci. U.S.A. 105 (2008) 14262-14264.

[9]  S. Margadonna, Y. Takabayashi, M. T. McDonald, K. Kasperkiewicz, Y. Mizuguchi, Y. Takano, A. N. Fitch, E. Suard, K. Prassides, Chem. Commun. (2008) 5607-5609.

[10]  A. Subedi, L. J. Zhang, D. J. Singh, M. H. Du, Phys. Rev. B 78 (2008) 134514.

[11]  Y. Mizuguchi, F. Tomioka, S. Tsuda, T. Yamaguchi, Y. Takano, App. Phys. Lett. 93 (2008) 152505.

[12]  D. Phelan, J. N. Millican, E. L. Thomas, J. B. Leão, Y. Qiu, R. Paul, Phys. Rev. B 79 (2009) 014519.

[13]  T. M. McQueen, Q. Huang, B. Ksenofontov, C. Felser, Q. Xu, H. Zandbergen, Y. S. Hor, J. Allred, A. J. Williams, D. Qu, J. Checkelsky, N. P. Ong, R. J. Cava, Phys. Rev. B 79 (2009) 014522.

[14]  http://www.ncnr.nist.gov/equipment/Pressure.html.

[15]  B. H. Toby, J. Appl. Cryst. 34 (2001) 210-213.

[16]  A. C. Larson; R. B. Von Dreele GSAS- Generalized Structure Analysis System; LANSCE, MS-H805, Los Alamos National Laboratory, Los Alamos, NM, 2000.

[17]  H. Takahashi, H. Okada, K. Igawa, K. Arii, Y. Kamihara, S. Matsuishi, M. Hirano, H. Hosono, K. Matsubayashi, Y. Uwatoko, J. Phys. Soc. Jpn. 77 (2008) 78-83.

[18]  J. Zhao, L. Wang, D. Dong, Z. Liu, H. Liu, G. Chen, D. Wu, J. Luo, N. Wang, Y. Yu, C. Jin, Q. Guo, J. Am. Chem. Soc. 130 (2008) 13828-13829.

[19]  A. Kreyssig, M. A. Green, Y. Lee, G. D. Samolyuk, P. Zajdel, J. W. Lynn, S. L. Bud'ko, M. S. Torikachvili, N. Ni, S. Nandi, J. B. Leao, S. J. Poulton, D. N. Argyriou, B. N. Harmon, R. J. McQueeney, P. C. Canfield, A. I. Goldman, Phys. Rev. B 78 (2008) 184517.

[20] J. Zhao, Q. Huang, C. de la Cruz, S. L. Li, J. W. Lynn, Y. Chen, M. A. Green, G. F. Chen, G. Li, Z. Li, J. L. Luo, N. L. Wang, P. C. Dai, Nature Mater. 7 (2008) 953-959.

[21] T. C. Ozawa, S. J. Kang, J. Appl. Cryst. 37 (2004) 679.

**Table 1. Refinement Results and Structural Parameters for FeSe.**
$T = 50$ K  Orthorhombic FeSe (*Cmme*)

| Pressure (GPa) | 0.028 | 0.204 | 0.391 | 0.602 |
|---|---|---|---|---|
| $a$ (Å) | 5.3088(2) | 5.3022(2) | 5.2959(3) | 5.2898(3) |
| $b$ (Å) | 5.3314(2) | 5.3245(2) | 5.3171(3) | 5.3116(3) |
| $c$ (Å) | 5.4865(1) | 5.4632(2) | 5.4442(2) | 5.4280(2) |
| $V$ (Å$^3$) | 155.124(13) | 154.234(9) | 153.306(10) | 152.512(11) |
| Fe ($u_{iso}$) | 0.0055(9) | 0.0050(6) | 0.0042(7) | 0.0010(1) |
| Se (position $z$) | 0.2659(6) | 0.2662(4) | 0.2679(4) | 0.2703(5) |
| Se ($u_{iso}$) | 0.0035(9) | 0.0039(7) | 0.0034(7) | 0.0066(9) |
| $\chi^2$ | 1.129 | 2.199 | 2.497 | 2.938 |
| $R_{wp}$ | 0.0944 | 0.0604 | 0.0614 | 0.0733 |
| $R_p$ | 0.0834 | 0.057 | 0.0584 | 0.0685 |

$T = 190$ K  Tetragonal FeSe (*P4/nmm*)

| Pressure (GPa) | 0.023 | 0.196 | 0.392 | 0.602 |
|---|---|---|---|---|
| $a$ (Å) | 3.7658(1) | 3.7610(1) | 3.7555(1) | 3.7501(1) |
| $c$ (Å) | 5.4988(2) | 5.4794(2) | 5.4598(2) | 5.4398(2) |
| $V$ (Å$^3$) | 77.980(5) | 77.505(5) | 77.008(5) | 76.503(5) |
| Fe ($u_{iso}$) | 0.0098(7) | 0.0091(7) | 0.0107(7) | 0.0088(7) |
| Se (position $z$) | 0.2663(4) | 0.2677(5) | 0.2688(4) | 0.2699(4) |
| Se ($u_{iso}$) | 0.0098(7) | 0.0086(8) | 0.0090(7) | 0.0076(8) |
| $\chi^2$ | 1.557 | 1.548 | 2.027 | 1.792 |
| $R_{wp}$ | 0.0679 | 0.0708 | 0.0685 | 0.0714 |
| $R_p$ | 0.0648 | 0.0661 | 0.0626 | 0.0660 |

**Table 2.** Isothermal Compressibility ($K$), Bulk Modulus ($B$), and Linear Compressibility ($K_a$, $K_b$, and $K_c$) of FeSe at $T = 50$ K and 190 K.

|  | 50 K | 190 K |
|---|---|---|
| $K$ (GPa$^{-1}$) | 0.031 | 0.033 |
| $B$ (GPa) | 33 | 31 |
| $K_a$ (10$^{-3}$ GPa$^{-1}$) | 6.3 | 7.2 |
| $K_b$ (10$^{-3}$ GPa$^{-1}$) | 6.9 | --- |
| $K_c$ (10$^{-3}$ GPa$^{-1}$) | 17.6 | 18.5 |

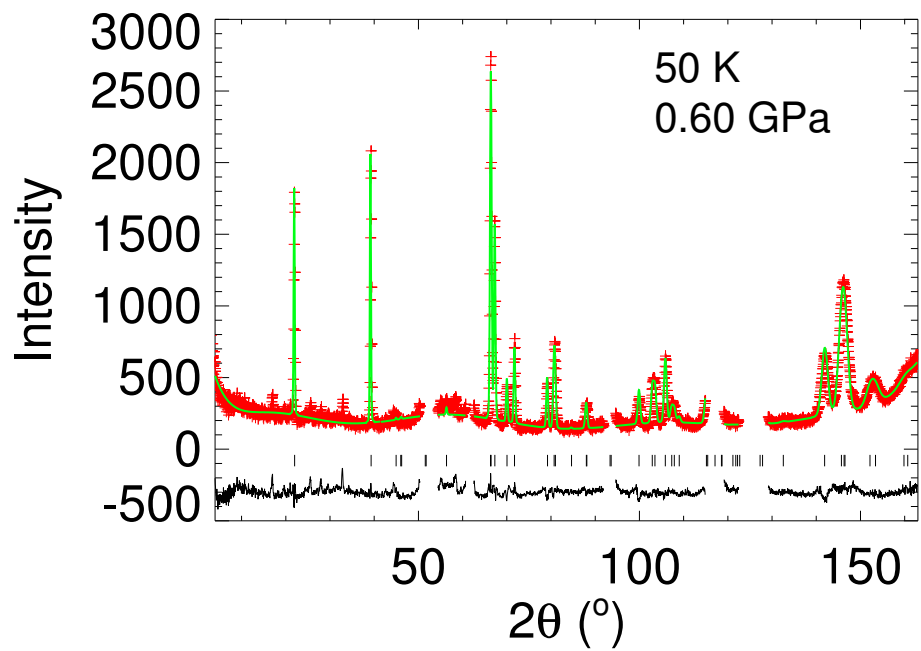

**Figure 1.**

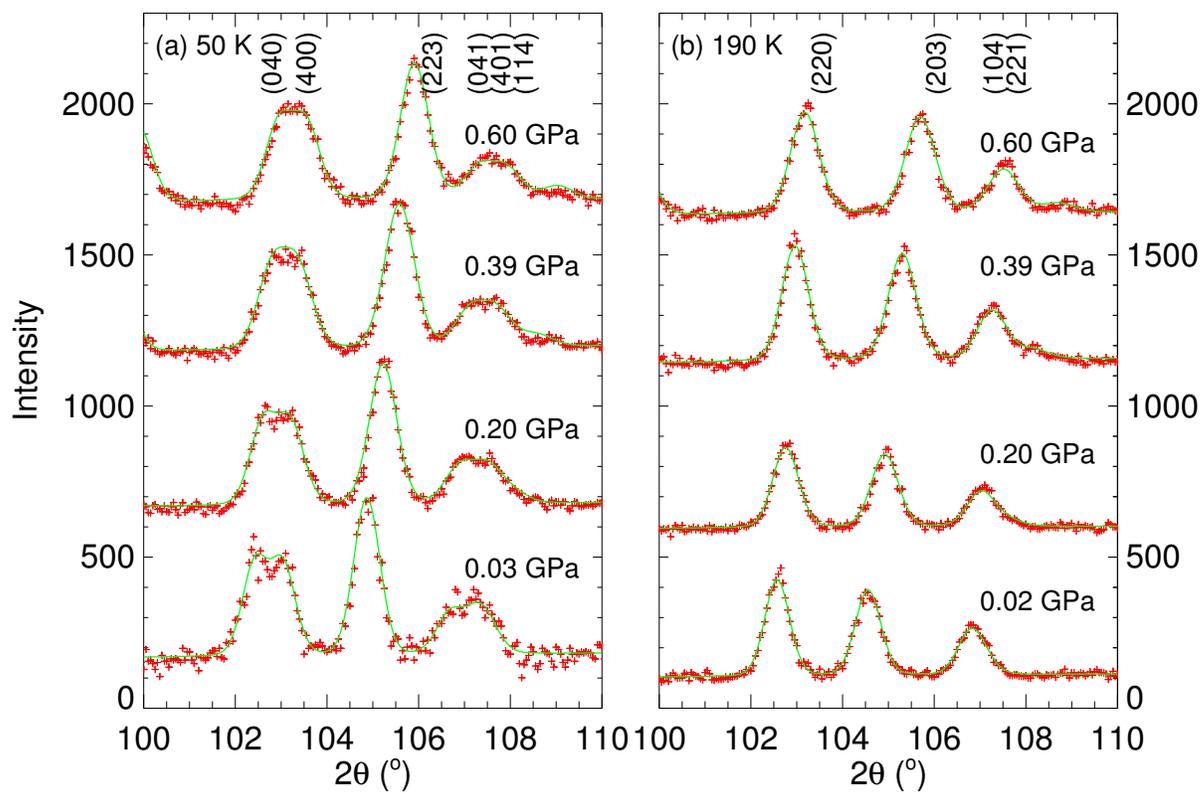

**Figure 2.**

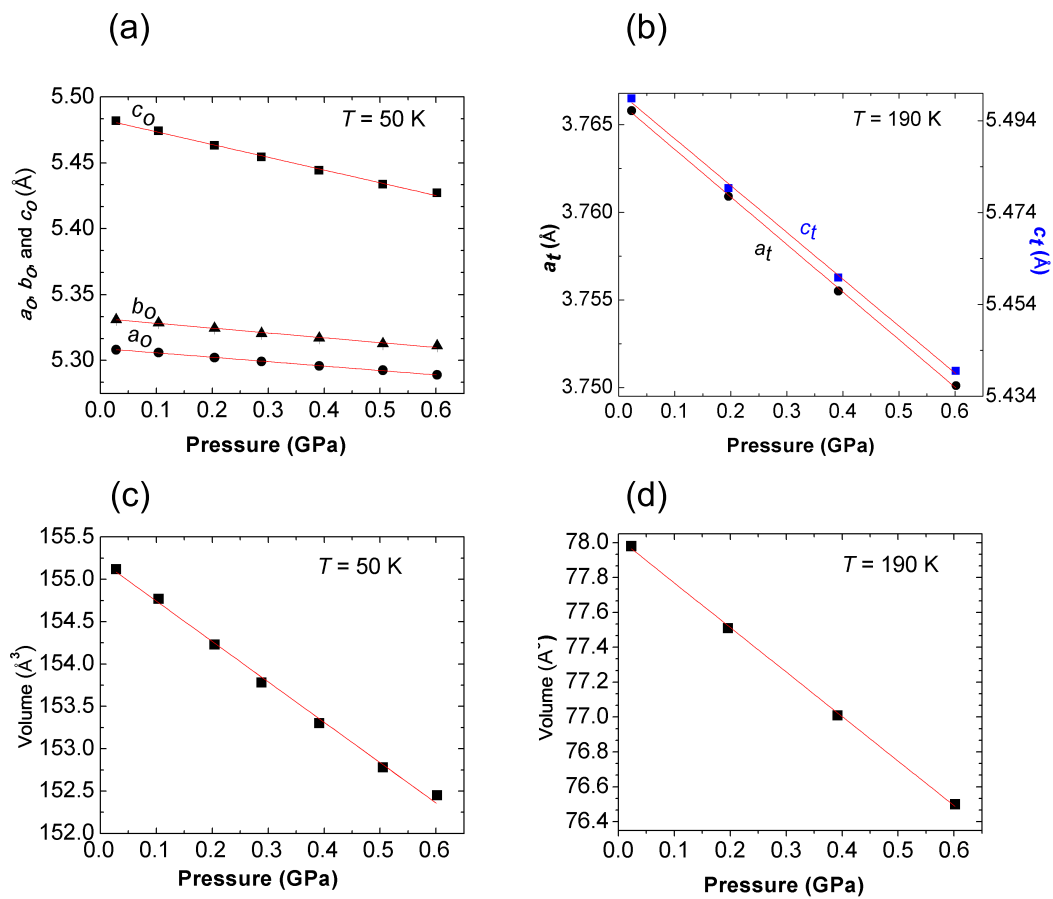

**Figure 3.**

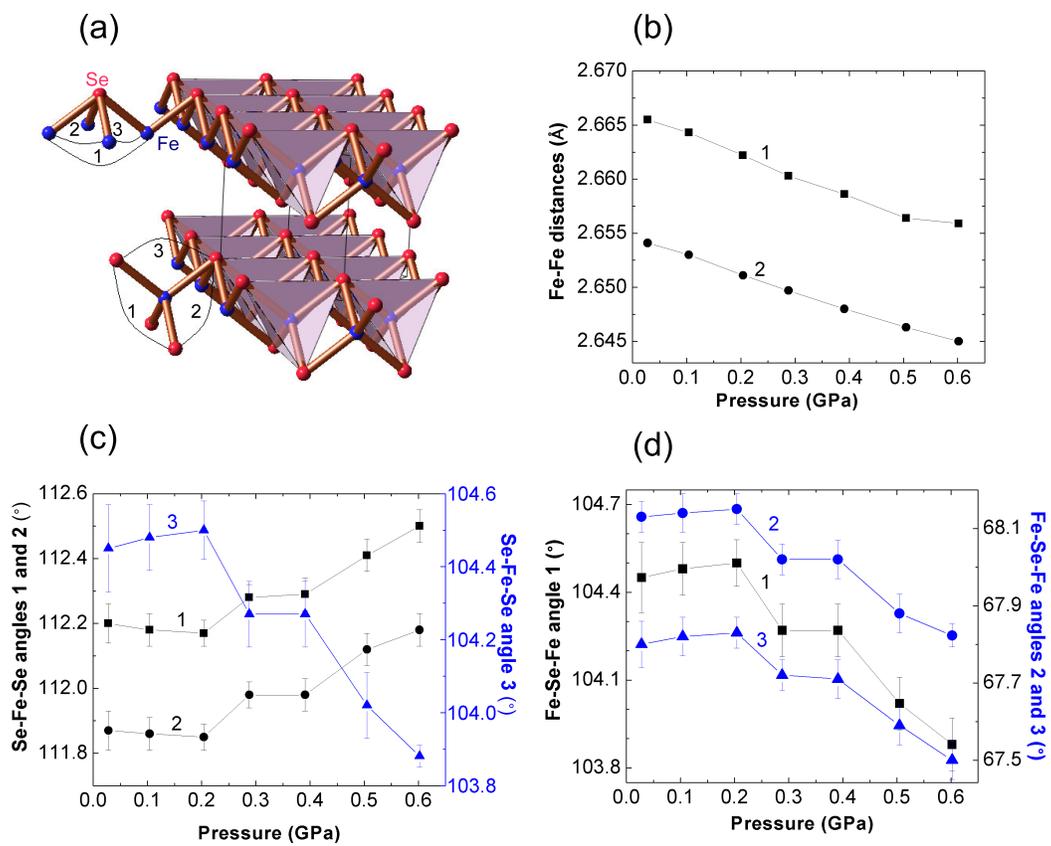

**Figure 4.**